%% file: main.tex
\documentclass[preprint]{vgtc}               

\usepackage{times}                     

\usepackage{enumitem} 

\graphicspath{{figs/}{figures/}{pictures/}{images/}{./}} 
\usepackage{tabu}                      
\usepackage{booktabs}                  
\usepackage{lipsum}                    
\usepackage{mwe}  
\usepackage{amsmath}
\usepackage{amsfonts}
\usepackage{mathptmx}    
\usepackage{makecell} 
\usepackage{graphicx} 
\usepackage{wrapfig}
\usepackage{fancyvrb}
\usepackage{mathptmx}                  

\vgtccategory{Research}

\vgtcinsertpkg

\title{When Do LLM Personas Support Visualization Design? A Cross-Model Study of Color Assignment and Chart Choice}

\author{Shahreen Salim\thanks{e-mail: ssalimaunti@cs.stonybrook.edu}\\ %
        \scriptsize Stony Brook University %
\and Klaus Mueller\thanks{e-mail: mueller@cs.stonybrook.edu}\\ %
     \parbox{1.4in}{\scriptsize \centering Stony Brook University}}

\abstract{

Large language model personas are increasingly used to approximate diverse users during early-stage visualization design, but it remains unclear whether persona-conditioned outputs reflect stable personality effects or artifacts of model choice and task framing. We examine this question across two visualization-relevant tasks: color assignment for abstract and concrete concepts, and chart-idiom preference ratings across task contexts. Using 43 Big Five profiles across GPT-4o-mini, GPT-4.1-mini, and GPT-5-mini, we find that personality-color coupling is highly model-configuration dependent: absent in GPT-4o-mini for all six concepts, consistent in GPT-4.1-mini across all six, and partial in GPT-5-mini for two of six. Concept type further shapes the signal: for abstract concepts, personality explains more hue variance than model identity, while concrete concepts show smaller and comparable effects. In chart choice, trait-aligned cluster aggregation produces stable top-idiom rankings across all nine cluster-context combinations, but a no-persona baseline recovers the same top choice in 8 of 9 model-context cells, indicating that task context drives rank-1 selection more than personality. These findings position LLM personas as exploratory probes for visualization design, not substitutes for human participants, and motivate multi-model testing, concept-type disaggregation, and no-persona baselines in future studies.

} 


\begin{document}


\firstsection{Introduction}

\maketitle

LLM personas are increasingly used to simulate diverse user profiles during early-stage visualization research~\cite{jiang2024evaluating, serapio2023personality}. The appeal is clear: if conditioning a model on a Big Five profile~\cite{John1999TheBF} produces systematic differences in visualization choices, researchers could use personas to explore design alternatives before running costly human studies. This use case is especially attractive for personalization problems, where color choices, chart preferences, and interpretation strategies may vary across users. However, the same convenience creates a methodological risk. Persona-conditioned outputs may reflect model-specific artifacts, prompt defaults, or task semantics rather than stable personality effects.

We examine this risk in two visualization-relevant settings. Experiment~1 tests whether Big Five persona conditioning changes the colors that LLMs assign to abstract and concrete concepts. Experiment~2 tests whether the same persona profiles produce stable chart-idiom preferences across task contexts, extending the chart-preference setting of Alves et al.~\cite{alves2020exploring}. Across both experiments, we compare GPT-4o-mini, GPT-4.1-mini, and GPT-5-mini using the same 43 unique Big Five profiles; Experiment~2 resamples these profiles to 60 persona-conditioned runs to approximate Alves et al.'s sample size.

Our results show that persona effects are not a single property of the prompt. In color assignment, personality-color coupling depends strongly on model configuration and concept type: the signal is absent in GPT-4o-mini, consistent in GPT-4.1-mini, and partial in GPT-5-mini, with stronger persona-driven variation for abstract concepts than concrete ones. In chart choice, trait-aligned cluster aggregation stabilizes rankings, but a no-persona baseline shows that rank-1 idiom selection is mostly context-driven. These findings support a methodological argument for persona-based visualization studies: test multiple model configurations, separate concept types, and report no-persona baselines before interpreting LLM persona outputs as preference signals. This is an exploratory study; we do not claim LLM personas substitute for human participants.

\input{sections/2.Related_Work.tex}
\input{sections/3.GPT_Personalities_and_Color}
\input{sections/4.GPT_personalities_and_Visualization_Preference}
\input{sections/7.Discussion}
\input{sections/8.Conclusion}

\section*{Supplemental Material}
All supplemental materials are available in the Supplemental.pdf. Supplemental materials include GPT prompts, profile sampling and trait-aligned cluster construction, Mantel binning and L1 vs.\ L2 distance sensitivity with exact per-concept statistics, circular-aware and per-model variance decomposition, multivariate trait-idiom regression coefficients, the full 12-idiom Borda heatmap and top-3 idioms by aggregation method, the cross-model Kendall's $\tau$ matrix, and the no-persona baseline.


\bibliographystyle{abbrv-doi}

\bibliography{template}
\end{document}

%% file: sections/2.Related_Work.tex
\section{Background and Related Work} \label{sec:relatedwork}

\paragraph{LLMs, color, and perceptual associations.}
LLMs encode associations between language and perceptual features such as color~\cite{mohammad2013colourful, Kawakami2016CharacterSM, abdou2021can, loyola2023perceptual}, and these associations extend beyond lexical co-occurrence~\cite{marjieh2023large}. Salim et al.~\cite{Salim2025Resonant} showed that LLMs generate concept-color associations that align with human semantic norms, using several of the same concepts we study here. This prior work motivates color assignment as a perceptual test case for persona-conditioned LLM behavior.

\paragraph{LLM personas and synthetic participants.}
LLMs can simulate personality-conditioned behavior in text~\cite{jiang2024evaluating, serapio2023personality, niszczota2023large}, though synthetic-participant work reports partial alignment rather than full behavioral equivalence~\cite{argyle2023one, dillion2023can, park2024generative}. Persona prompting also introduces clear risks: models can fall back on demographic caricatures~\cite{Shu2023YouDN, cheng2023marked}, inherit reasoning biases~\cite{gupta2024bias}, and behave inconsistently under formal assessment~\cite{song2023large}. Work on LLM-as-a-judge evaluation~\cite{zheng2023judging} further shows that LLM preferences can reflect training artifacts rather than generalizable principles. We therefore do not treat these systems as having intrinsic personalities; we test whether prompting distinct Big Five profiles produces systematic output differences that persist across model configurations.

\paragraph{Personality and visualization preference.}
Prior work links personality to visualization preference. Alves et al.~\cite{alves2020exploring} found that Big Five traits correlate with chart preferences across contexts, although their study used only $N=64$ participants. This connects to visualization recommendation and chart-preference work, where task semantics often constrain chart choice before individual preferences modulate the final design. We extend this line of work by testing whether persona conditioning produces variation in both perceptual outputs (color choices) and higher-level comparative judgments (chart preferences), and when model effects or concept type may override the persona signal.

%% file: sections/3.GPT_Personalities_and_Color.tex
\section{Experiment 1: GPT and Personality Traits in Color Generation} \label{sec: case-study1}

We test whether Big Five personality conditioning~\cite{John1999TheBF} changes the colors that LLM personas assign to concepts across GPT-4o-mini, GPT-4.1-mini, and GPT-5-mini.

\subsection{Experimental Settings}

We selected six concepts spanning two concept types, drawn from Salim et al.~\cite{Salim2025Resonant}. The concrete set contains \textit{Banana}, \textit{Strawberry}, and \textit{Carrot}, which have strong default color associations. The abstract set contains \textit{Serendipity}, \textit{Serenity}, and \textit{Chaos}, which allow wider interpretation. This contrast lets us test whether personality-driven variation depends on concept type.

\subsubsection{Big Five Trait Profiles} \label{uspopulation-43}

We use 43 profiles derived from Rentfrow et al.'s state-level U.S. personality dataset~\cite{Rentfrow2013PERSONALITYPA}. We converted each state's Big Five T-score into Low, Average, or High categorical levels and removed duplicate trait combinations across states, yielding 43 unique profiles. Experiment~2 resamples these profiles to 60 persona-conditioned runs to approximate Alves et al.'s sample size; duplicated profiles should be interpreted as repeated stochastic runs, not as 60 independent personality profiles. Prompt details appear in the supplement.

\begin{figure}[h]
     \centering
\includegraphics[width=\columnwidth]{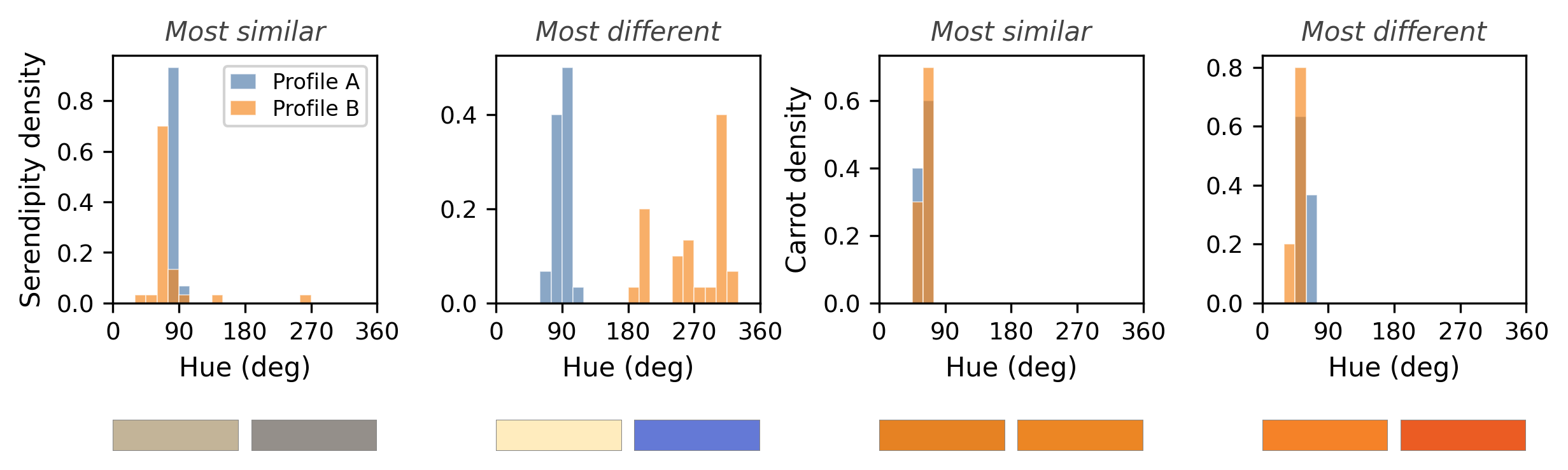}
 \vspace{-6pt}
   \caption{GPT-4.1-mini hue distributions for the most-similar (top) and most-different (bottom) persona pairs on Serendipity (abstract) and Carrot (concrete). Most-similar: O=Low/C=Avg/E=Avg/A=Avg/N=Avg vs.\ N=High. Most-different: O=Low/C=High/E=High/A=High/N=Low vs.\ O=High/C=Low/E=Low/A=Low/N=High.}
     \vspace{-10pt}
     \label{fig:profile_similarity-lch-distribution}
\end{figure}

\subsubsection{Color Generation from Concepts and Personality}

For each concept-persona pair, we collected 30 colors per model. We prompted each model to return a Python list of dictionaries in the form \texttt{\{`name', `r', `g', `b'\}} with integer RGB values in [0, 255] (full prompt in supplement). We parsed responses with a permissive JSON/Python-literal reader, rejected malformed or out-of-range outputs, and reissued calls until each concept-persona-model cell contained 30 valid samples. Each sRGB triple was converted to CIELCH through the standard sRGB $\rightarrow$ linear RGB $\rightarrow$ CIEXYZ (D65) $\rightarrow$ CIELAB $\rightarrow$ CIELCH pipeline. Each persona's hue distribution was then represented as a 36-bin circular histogram over hue angle $H$ (10$^{\circ}$ per bin).

GPT-4o-mini and GPT-4.1-mini were queried with temperature $= 0.7$. GPT-5-mini does not expose temperature, so we induced variability across 30 fixed seeds using eight interpretive lenses (literal, affective, metaphorical, atmospheric, muted, vivid, dark, light). Because the same seeds and lens assignments were used for every persona within a concept, lens-induced variance is shared across personas and cannot by itself create between-persona structure in the pairwise distance matrix; any personality-color coupling must therefore come from the persona prompt. This protocol differs from temperature-based sampling, and we cannot fully rule out that lens assignment interacts with concept type in ways that do not generalize to a temperature-equivalent protocol; future work should test GPT-5-series models with any API-level randomization controls that become available.

We summarize between-persona variation using average pairwise Hellinger distance~\cite{beran1977hellinger} across the 43 persona-specific hue histograms. On GPT-4.1-mini, abstract concepts show both higher within-persona dispersion (CIEDE2000 between each sample and its profile mean: 12.7--31.1) and higher between-persona Hellinger distance (0.57--0.79) than concrete concepts (CIEDE2000: 5.6--9.6; Hellinger: 0.37--0.41), indicating broader and more persona-sensitive hue distributions. Per-concept LCH density across all 43 profiles appears in Supplementary Figure~S2.

\subsection{Personality Distance vs. Color Distance}
\label{sec:mantel}

We test whether personas that are farther apart in Big Five space also produce more different hue distributions. For each model, we construct a personality-distance matrix using L1 distance over five ordinal trait levels (Low $= 0$, Average $= 1$, High $= 2$). We also construct a color-distance matrix using Hellinger distance over persona-specific hue histograms (36 bins, 10$^{\circ}$ per bin in CIELCH $H$). Because each persona appears in 42 pairwise comparisons, these distances are not independent; we therefore use a Mantel permutation test~\cite{mantel1967detection} with 9,999 permutations. We convert Mantel $r$ to Cohen's $d$ using $d = 2r/\sqrt{1-r^2}$.

Personality-color coupling differs sharply across model configurations. At 36 hue bins, GPT-4o-mini shows no significant association for any concept (0/6; all $p > 0.16$). GPT-4.1-mini shows consistent coupling across all six concepts (6/6; $p \leq 0.006$, $r = 0.12$--$0.21$). GPT-5-mini shows partial coupling, with significant associations only for Strawberry ($p = 0.001$) and Chaos ($p = 0.015$). After Holm correction across all 18 tests, four GPT-4.1-mini concepts and GPT-5-mini Strawberry remain significant. Exact $r$ values, raw and Holm-corrected $p$-values, and converted Cohen's $d$ values appear in Supplementary Table~S2.

This pattern is stable for GPT-4o-mini and GPT-4.1-mini but less stable for GPT-5-mini. Re-running the Mantel test at 18, 36, and 72 hue bins yields the same concept count for GPT-4o-mini (0/6 at all resolutions) and GPT-4.1-mini (6/6 at all resolutions). GPT-5-mini varies across resolutions (4/6, 2/6, 2/6), indicating that its weaker effects are more sensitive to hue-bin resolution. Supplementary Table~S1 summarizes this binning-sensitivity analysis.
\Cref{fig:profile_similarity-lch-distribution} illustrates this pattern for GPT-4.1-mini; per-concept LCH density across all 43 profiles appears in Supplementary Figure~S2.

\subsection{Cross-Model Comparison}
\label{sec:cross-model}

\begin{figure}[ht]
\centering
\includegraphics[width=\columnwidth]{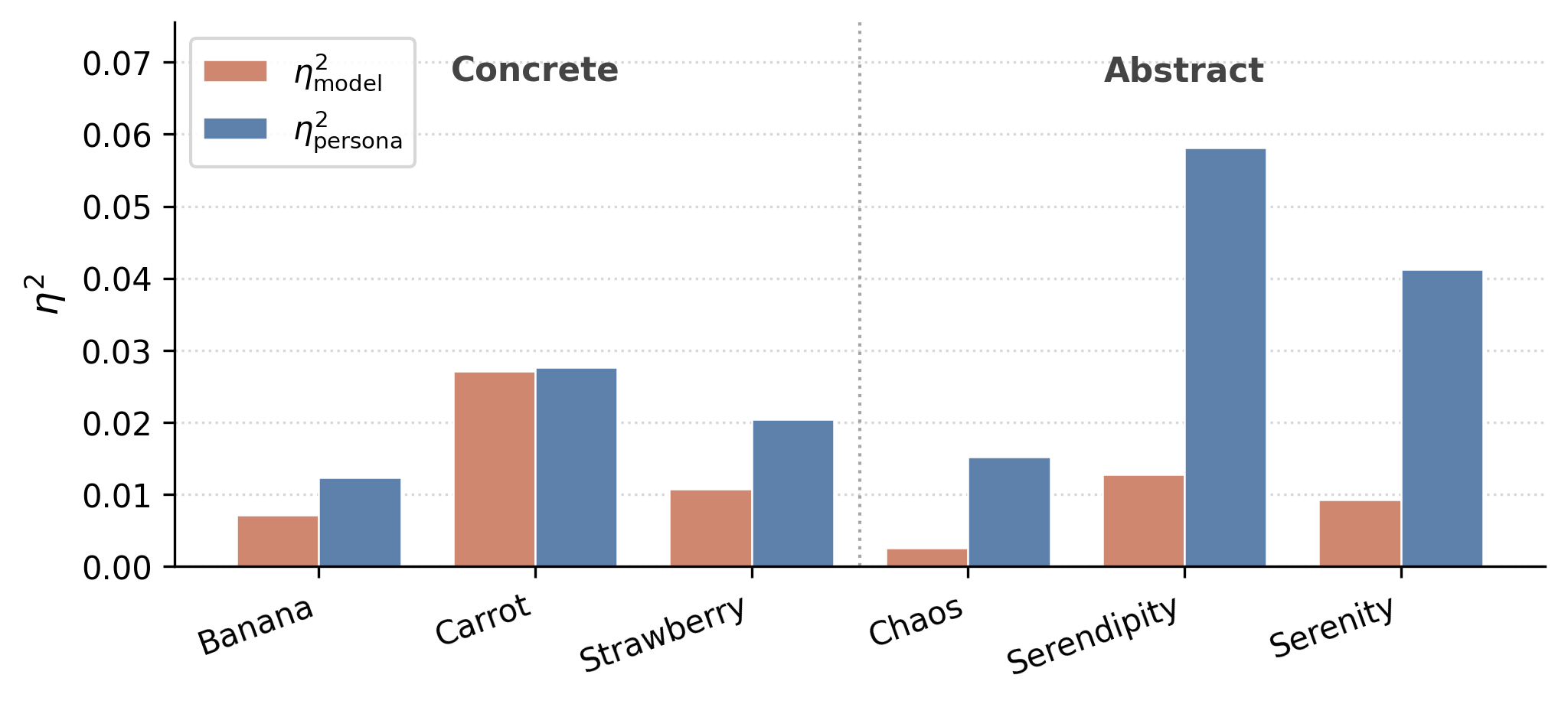}
\caption{Variance decomposition of hue (H\textsubscript{LCH}) from a two-way ANOVA (model $\times$ persona) pooled across all three models. Persona $\eta^2$ exceeds model $\eta^2$ for all abstract concepts.}
\label{fig:variance}
\vspace{-6pt}
\end{figure}

We next compare the relative contribution of model identity and persona to hue variance. As a descriptive variance decomposition, we fit a two-way ANOVA on H\textsubscript{LCH} with model and persona as factors and report $\eta^2$ for each source (Figure~\ref{fig:variance}). Because samples are nested within personas, models, concepts, and sampling protocols, we treat this decomposition as an effect-size summary rather than an inferential test.

Figure~\ref{fig:variance} shows that concept type changes the balance between model and persona effects. For concrete concepts, both effects are small and similar on average ($\eta^2_{\text{model}} = 0.015$, $\eta^2_{\text{persona}} = 0.020$), while residual variance dominates each fit (0.929--0.982). For abstract concepts, persona explains more hue variance than model identity ($\eta^2_{\text{persona}} = 0.038$ vs.\ $\eta^2_{\text{model}} = 0.008$), a 4.7$\times$ difference. This indicates that abstract concepts leave more room for persona-conditioned variation, whereas concrete concepts are more constrained by default color associations.

Because hue is circular, we replicated the decomposition on $\sin H$ and $\cos H$ and averaged $\eta^2$ across components (Supplementary Table~S3). The abstract-concept ordering remains the same ($\bar{\eta}^2_{\text{persona}} = 0.036$ vs.\ $\bar{\eta}^2_{\text{model}} = 0.018$), although the gap narrows from 4.7$\times$ to 2.0$\times$. Concrete concepts are more sensitive to this circular treatment: persona $>$ model holds for Strawberry and Carrot, but reverses for Banana.

The pooled abstract-concept effect is mainly a GPT-4.1-mini effect. Per-model within-model $\eta^2_{\text{persona}}$ (Supplementary Table~S4) is much larger for GPT-4.1-mini (mean $\eta^2_{\text{persona}} = 0.23$) than for GPT-4o-mini (0.05) or GPT-5-mini (0.03). Thus, the larger structured hue effect comes from persona variation within GPT-4.1-mini, not from model identity alone.

Together, the Mantel test and variance decomposition show that personality-color coupling is configuration-dependent and stronger for abstract concepts. However, residual variance dominates each per-concept fit, so per-persona color output remains noisy. Experiment~2 therefore tests whether a coarser unit of analysis, trait-aligned persona clusters, yields more stable design-preference signals.

%% file: sections/4.GPT_personalities_and_Visualization_Preference.tex
\section{Experiment 2: LLM Personas and Chart Preferences}

Experiment~1 shows that persona-conditioned color signals are fragile at the individual-profile level and configuration-dependent. We now test whether higher-level design preferences become more stable after aggregating personas into trait-aligned clusters. Building on Alves et al.'s chart-preference task~\cite{alves2020exploring} ($N=64$ human participants), Experiment~2 asks whether persona conditioning changes chart-idiom rankings beyond context-driven defaults, or mainly modulates ratings within a context-fixed ranking.

\subsection{Dataset Generation and Simulation}

We use the 12 chart idioms and three task contexts from Alves et al.: hierarchy, time series, and comparison. We recreated each chart in high resolution with matched axes, data, and palettes. For each context, LLM personas answered prompts modeled on Alves's survey (for example, ``Which chart best represents a time-evolving dataset?'') and rated each idiom on Clarity, Interpretability, Appeal, and Overall Preference using a 7-point scale. Chart order and prompt phrasing were held fixed across all personas within each context; position-order effects, if present, would appear equally across clusters and would not explain between-cluster differences, but we cannot rule them out as a source of absolute-level bias.

Experiment~2 uses 60 persona-conditioned runs drawn from the 43 unique Big Five profiles in Experiment~1, with some profiles sampled multiple times to approximate Alves et al.'s human sample size ($N = 64$). We assign each persona to one of three Alves-style trait prototypes by L1 distance over the 5-D ordinal trait vector (Low $= 0$, Average $= 1$, High $= 2$): Cluster~1 \textit{Organized and Stable} (High C, Low N; $n=24$), Cluster~2 \textit{Sociable and Cooperative} (High E, High A, Low C; $n=14$), Cluster~3 \textit{Emotionally Reactive} (High N, Average C; $n=22$). Seven personas tie between two prototypes and are assigned to the lower-index prototype. Cluster trait means recover the prototype directions (full assignment in supplement).

\subsection{Trait--Preference Associations}
\label{sec:trait-correlation}

We treat trait-aligned cluster aggregation as the primary Experiment~2 analysis. We first test individual-trait associations using one multivariate OLS regression per idiom on GPT-4.1-mini ratings (12 fits). Each model includes all five Big Five traits jointly: $\text{overall\_rating} \sim \text{O} + \text{C} + \text{E} + \text{A} + \text{N}$, where Low $= 0$, Average $= 1$, and High $= 2$. This joint fit avoids aliasing correlated traits onto one another. Each idiom-level fit uses one rating from each of the 60 persona-conditioned runs (df\textsubscript{resid} $= 54$). Under Holm-Bonferroni correction across all 60 trait-idiom coefficients at $\alpha = 0.05$, 36 coefficients remain significant, and all 60 coefficient signs match their univariate counterparts. Full multivariate results appear in Supplementary Figure~S3 and Table~S5. These associations are exploratory: 60 runs per fit, no held-out validation, and regressions fit on GPT-4.1-mini only.

\subsection{Cluster-Based Preference Analysis}
\label{sec:cluster-preference}

We next analyze preferences at the trait-aligned cluster level using four aggregation methods: Majority Vote (plurality of rank-1 idioms), IRV (Instant Runoff Voting), Apriori Support (fraction of personas in which an idiom appears in the top-3), and Borda Count~\cite{borda1781memoire} (inverse-rank weighted). All percentages below are Borda shares within context.

\begin{figure*}[h]
  \centering
  \begin{minipage}[c]{0.53\linewidth}
    \centering
    \includegraphics[width=\linewidth]{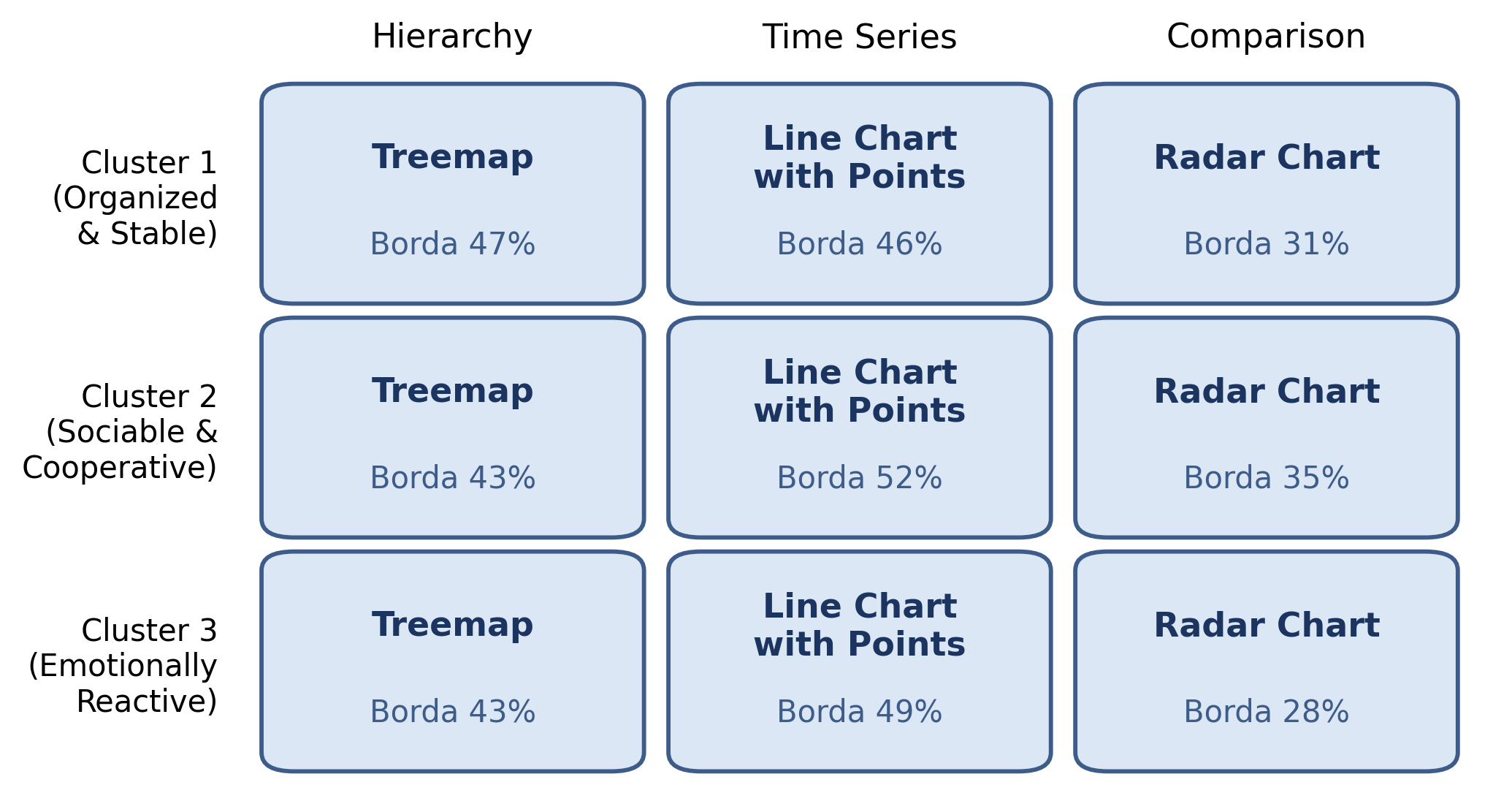}\\
    {\footnotesize (a) Borda share within context}
  \end{minipage}\hfill%
  \begin{minipage}[c]{0.44\linewidth}
    \centering
    \includegraphics[width=\linewidth]{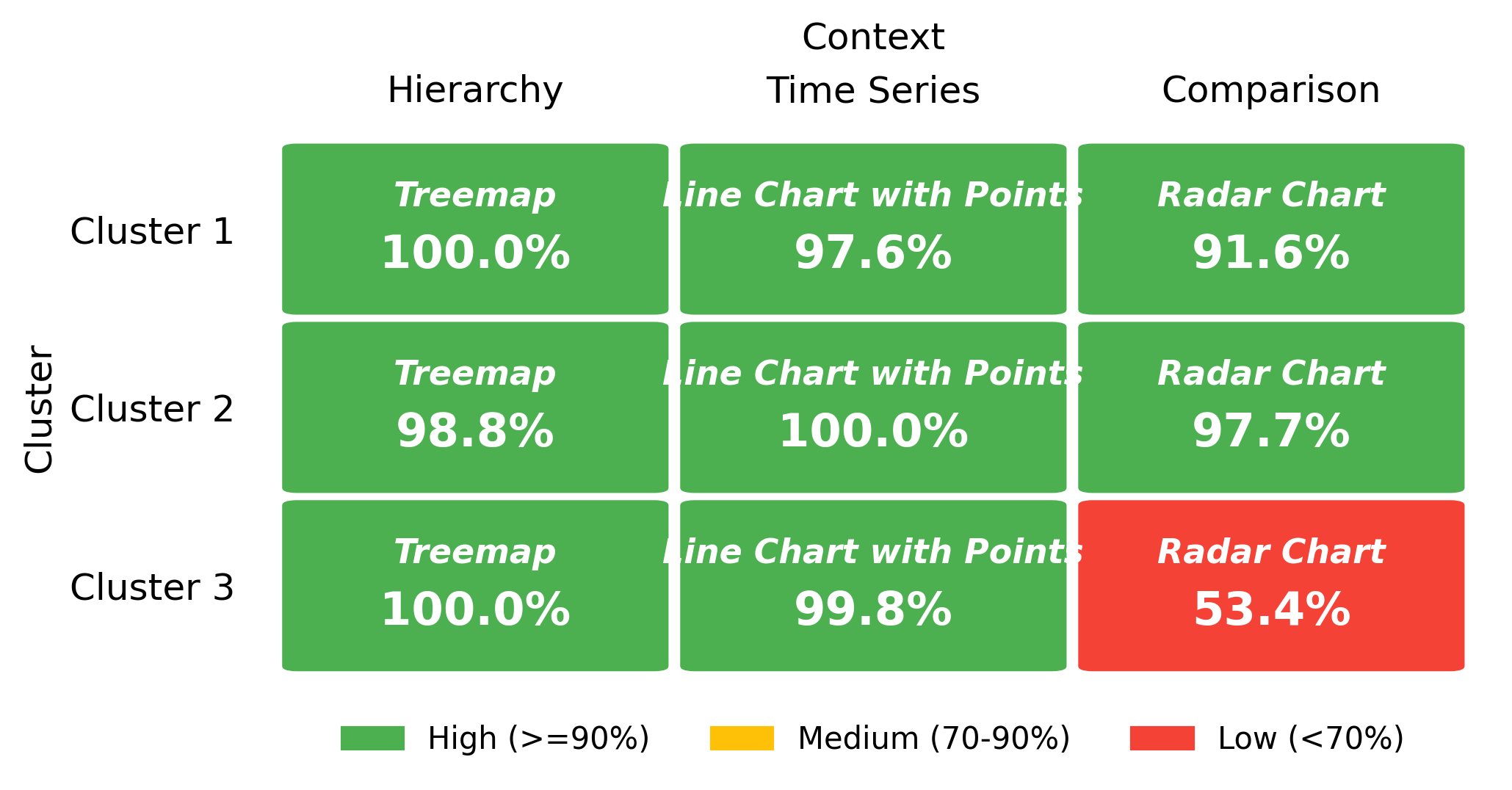}\\
    {\footnotesize (b) Bootstrap stability of rank-1 idiom}
  \end{minipage}
\caption{Cluster-based idiom preferences on GPT-4.1-mini. (a) Borda rank-1 idiom and within-context share per (cluster, context). (b) Borda rank-1 bootstrap stability (\%) per (cluster, context) (1{,}000 persona-level resamples). Comparison-context stability drops to 53\% for Cluster~3 (\textit{Emotionally Reactive}, $n{=}22$). The full 12-idiom Borda share for GPT-4.1-mini is shown in Supplementary Figure~S4; full top-3 rankings per (model, cluster, context) cell appear in Supplementary Table~S6.}
  \label{fig:cluster_weighted_heatmap}
\end{figure*}

Figure~\ref{fig:cluster_weighted_heatmap} shows context-dominant agreement. All three trait-aligned clusters rank Treemap first for hierarchy (Borda shares 47\%, 43\%, 43\%), Line Chart with Points first for time series (46\%, 52\%, 49\%), and Radar Chart first for comparison (31\%, 35\%, 28\%) under Majority Vote, IRV, and Borda. These choices diverge from Alves et al.~\cite{alves2020exploring} in hierarchy (Sunburst) and time series (line charts without points), so we treat Alves et al. as a task framework and reference point, not as a direct replication target.

We estimate ranking stability with 1{,}000 persona-level bootstrap resamples under Borda aggregation (seed = 42). Hierarchy and time-series top choices are nearly perfectly stable across all three clusters (97.6--100\%; Figure~\ref{fig:cluster_weighted_heatmap}, right panel). The comparison context is less stable: Cluster~1 remains high at 91.6\%, Cluster~2 at 97.7\%, but Cluster~3 drops to 53.4\%. We therefore treat the Cluster~3 comparison rank-1 result as suggestive rather than settled.

Apriori Support diverges from Majority Vote, IRV, and Borda in 5 of 27 cluster-context-model cells, mainly in the comparison context where it returns Pie Chart. However, top-3 sets agree even when rank-1 differs (Supplementary Table~S6). An alternative mean-Likert aggregation also yields different top idioms and lower stability (supplement), reinforcing our use of rank-based aggregation for the main analysis.

We also control for visual complexity. For each model and context, we fit a covariate-adjusted mixed-effects model: $\text{overall\_rating} \sim \text{cluster} + \text{min\_colors\_required} + (1|\text{persona\_id})$. No top-ranked idiom changes under this control in any of the 27 cluster-context-model cells. Pooling across context for GPT-4.1-mini, cluster intercepts (Cluster~1 as reference) are substantial and significant after Holm correction: Cluster~2 vs.\ Cluster~1: $\beta = -0.44$, $SE = 0.13$, $t = -3.39$, $p_{\text{Holm}} = 0.0007$; Cluster~3 vs.\ Cluster~1: $\beta = -0.93$, $SE = 0.12$, $t = -8.05$, $p_{\text{Holm}} < 0.0001$. The \textit{Emotionally Reactive} cluster (high N) rates idioms about one rating point lower than the \textit{Organized and Stable} cluster on a 7-point scale, so cluster-level differences appear in absolute rating level and lower-ranked ordering, even when the rank-1 idiom remains unchanged.

Together, these results show that trait-aligned aggregation stabilizes top-idiom rankings for hierarchy and time-series contexts across model configurations, while the comparison context remains less stable for the \textit{Emotionally Reactive} cluster.

\subsection{Cross-Model Comparison}
\label{sec:exp2-cross-model}

We finally repeat the cluster-based preference analysis across GPT-4o-mini, GPT-4.1-mini, and GPT-5-mini using identical prompts and the same persona profiles. For each of the nine cluster-context pairs, we compute Kendall's $\tau$ between each pair of models over the full 12-idiom Borda ranking.

Cross-model agreement is strongest at the top-choice level and weaker for full 12-idiom rankings. All nine cluster-context pairs show unanimous top-idiom agreement across all three models under Majority Vote, IRV, and Borda: Treemap in hierarchy, Line Chart with Points in time series, and Radar Chart in comparison, for every cluster under every model. A no-persona baseline recovers the same top idiom in 8 of 9 model-context cells, indicating that this cross-model convergence largely reflects context-driven defaults rather than persona-specific top-choice reordering.

Mean Kendall's $\tau$ across the 27 cluster-context-model-pair cells is 0.79; averaged within each model pair across the 9 cluster-context cells, $\bar{\tau}$(4o-mini, 4.1-mini) $= 0.889$, $\bar{\tau}$(4o-mini, 5-mini) $= 0.778$, $\bar{\tau}$(4.1-mini, 5-mini) $= 0.711$. The 5 of 27 cluster-context-model pairs with $\tau < 0.5$ are concentrated in the time-series and comparison contexts (4 in Cluster~3, 1 in Cluster~2 comparison), consistent with the lower bootstrap stability in those cells. Exact $\tau$ values per cluster-context-model pair appear in Supplementary Table~S7. Figure~\ref{fig:cross_model_heatmap} summarizes the agreement pattern.

\begin{figure}[h]
  \centering
  \includegraphics[width=\linewidth]{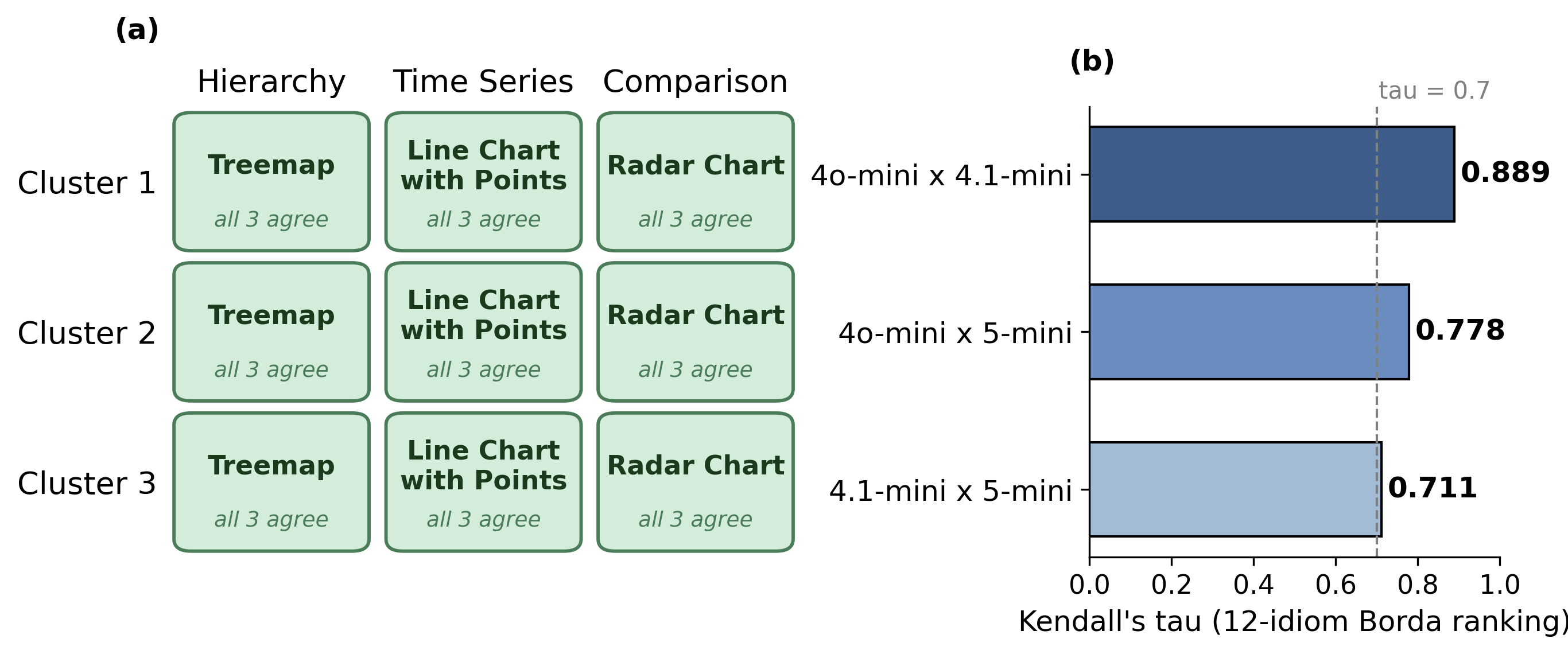}
  \caption{Cross-model agreement on chart preference. (a) Top-ranked idiom is unanimous across all three models for every trait-aligned cluster-context cell: Treemap (hierarchy), Line Chart with Points (time series), Radar Chart (comparison). (b) Pairwise Kendall's $\tau$ on 12-idiom Borda rankings; the 4.1-mini vs.\ 5-mini pair is lowest.}
  \label{fig:cross_model_heatmap}
  \vspace{-6pt}
\end{figure}

%% file: sections/7.Discussion.tex
\section{Discussion}
\label{sec:discussion}

Our results support using LLM persona simulation as an exploratory method for probing visualization design choices, not as a substitute for human participants. Across both experiments, the central pattern is consistent: persona conditioning can produce measurable structure, but that structure depends on model configuration, task type, and analysis granularity. These dependencies shape how persona-based visualization studies should be interpreted and reported.

\textbf{Persona effects are model-configuration dependent.}
The color experiment shows that personality-color coupling is not a stable property of persona prompting alone. Under a Mantel permutation test that respects the dependence structure of pairwise comparisons, coupling is absent in GPT-4o-mini, consistent in GPT-4.1-mini, and partial in GPT-5-mini. GPT-4o-mini still produces variable color outputs across personas; what disappears is structured alignment between personality distance and hue-distribution distance. This distinction matters because visual variability alone can be mistaken for personality signal. Since GPT-5-mini required a lens-based variability protocol, its result should be read as a stress test rather than a temperature-matched generation comparison. Multi-model testing is therefore necessary for persona-based visualization studies.

\textbf{Concept type changes where persona signal appears.}
Abstract concepts leave more room for persona-conditioned interpretation than concrete concepts. For abstract concepts, personality explains more hue variance ($\eta^2 = 0.038$) than model identity ($\eta^2 = 0.008$). For concrete concepts, both contributions are small and within one percentage point ($\eta^2_{\text{model}} = 0.015$, $\eta^2_{\text{persona}} = 0.020$). This supports the interpretation that concrete concepts are constrained by strong default color associations, while abstract concepts allow more interpretive flexibility. Pooling concept types would therefore hide both the abstract-concept persona signal and the concrete-concept null.

\textbf{Aggregation stabilizes rankings, but does not imply personality-driven top choices.}
In Experiment~2, trait-aligned cluster aggregation produces stable chart-idiom rankings for hierarchy and time-series contexts. Majority Vote, IRV, and Borda Count converge on the same top idiom within each cluster, and all nine cluster-context combinations show the same top idiom across all three models. Apriori Support diverges in 5 cells, mostly in the comparison context, indicating that rank-1 stability is weaker when top-3 support is considered. These results show that cluster-level aggregation is more stable than individual-trait fits for exploring relative preference ordering. However, the multivariate trait regressions in \S\ref{sec:trait-correlation} remain exploratory and should not be interpreted as causal personality effects.

\textbf{Task context dominates chart top-choice selection.}
The agreement across trait-aligned clusters and models indicates that context semantics (hierarchy, time series, comparison) drive the rank-1 chart choice more strongly than persona conditioning. A no-persona baseline replicates the persona-conditioned top idiom in 8 of 9 (model, context) cells; the one disagreement, GPT-5-mini comparison, is a near-tie in which Radar Chart is the no-persona rank-2. Persona prompting therefore contributes rating-level modulation rather than top-idiom reordering. For example, the \textit{Emotionally Reactive} cluster rates idioms about one point lower than the \textit{Organized and Stable} cluster on the 7-point scale (mixed-effects $\beta = -0.93$, $p_{\text{Holm}} < 0.0001$), but this shift does not change the rank-1 idiom in any context.

\textbf{Limitations and future work.}
The absence of matched human validation is both a limitation and a deliberate scope choice. We characterize out-of-the-box persona-conditioned LLM behavior before introducing human-in-the-loop calibration, which would change the object of study from raw persona prompting to a tuned system. Matched human calibration is the next step for determining which observed patterns reflect human preference and which are model artifacts. The six-concept set is sufficient for a methodological demonstration but too small to support broad generalization. Mantel results are stable under L1 and Euclidean distance over the same ordinal encoding (supplement), but future work should test psychologically motivated trait-distance functions. Seven of 60 personas tie between prototypes; excluding them preserves the rank-1 idiom in all nine cluster-context cells across all three models. GPT-5-mini's lens protocol, cross-cultural and cross-language generalization, and persona-prompting bias remain open limitations~\cite{cheng2023marked, gupta2024bias}.

%% file: sections/8.Conclusion.tex
\section{Conclusion}
\label{sec:conclusion}

LLM personas can expose structured variation in visualization design tasks, but that variation is not stable enough to treat persona prompting as a direct proxy for user preference. In color assignment, personality-color coupling differs sharply across model configurations and is strongest for abstract concepts. In chart choice, trait-aligned aggregation stabilizes rankings, but the no-persona baseline shows that top idioms are mostly context-driven; persona prompting mainly modulates rating levels rather than reordering the top choice. These results support LLM personas as exploratory probes for visualization design, not substitutes for human participants. Stronger claims require multi-configuration evaluation, concept-type disaggregation, no-persona baselines, and small-scale human calibration.